\documentclass[superscriptaddress,amssymb,aps,pra,showpacs,footinbib,twocolumn]{revtex4-1}
\usepackage{amsmath}
\usepackage{float}

\usepackage{graphicx}
\usepackage{sidecap}
\usepackage{dcolumn}
\usepackage{bm}
\usepackage{hyperref}
\usepackage{subfigure}
\usepackage{verbatim}
\graphicspath{ {./pictures/} }
\usepackage{color}

\begin{document}

\title{Controlled polarization of two-dimensional quantum turbulence 
in atomic Bose-Einstein condensates}

\author{A. Cidrim}
\affiliation{Instituto de F\'{i}sica de S\~{a}o Carlos, Universidade de S\~{a}o Paulo, Caixa Postal 369, 13560-970 S\~{a}o Carlos, S\~{a}o Paulo, Brazil}
\affiliation{JQC (Joint Quantum Centre Durham-Newcastle) and
School of Mathematics and Statistics, Newcastle University, Newcastle upon Tyne, NE1 7RU, United Kingdom}

\author{F. E. A. dos Santos}
\affiliation{Departamento de F\'{i}sica, Universidade Federal de S\~{a}o Carlos, 13565-905, S\~{a}o Carlos, SP, Brazil}

\author{L. Galantucci}
\affiliation{JQC (Joint Quantum Centre Durham-Newcastle) and
School of Mathematics and Statistics, Newcastle University, Newcastle upon Tyne, NE1 7RU, United Kingdom}

\author{V. S. Bagnato}
\affiliation{Instituto de F\'{i}sica de S\~{a}o Carlos, Universidade de S\~{a}o Paulo, Caixa Postal 369, 13560-970 S\~{a}o Carlos, S\~{a}o Paulo, Brazil}

\author{C. F. Barenghi}
\affiliation{JQC (Joint Quantum Centre Durham-Newcastle) and
School of Mathematics and Statistics, Newcastle University, Newcastle upon Tyne, NE1 7RU, United Kingdom}


\begin{abstract}

We propose a scheme for generating two-dimensional
turbulence in harmonically trapped atomic condensates
with the novelty of controlling the polarization 
(net rotation) of the
turbulence. Our scheme is based on an initial giant
(multicharged) vortex which induces a large-scale circular flow.
Two thin obstacles, created by blue-detuned
laser beams, speed up the decay of the
giant vortex into many singly-quantized vortices of the
same circulation; at the same time, vortex-antivortex pairs are
created by the decaying circular flow past the obstacles.
Rotation of the obstacles against the circular flow
controls the relative proportion of
positive and negative vortices, from the limit of strongly
anisotropic turbulence (almost all vortices having the same sign) to
that of isotropic turbulence (equal number of vortices and
antivortices). Using the new scheme, we numerically study 
quantum turbulence and report on its decay
as a function of the polarization. We finally present a phenomenological model for the decay rate of vortex number which fits our numerical experiment curves, with the novelty of taking into account polarization time-dependence.
\end{abstract}

\maketitle

\section{Introduction}

The study of quantum turbulence is heavily 
motivated by liquid helium ($^4$He and $^3$He) experiments 
\cite{Skrbek2012,Barenghi2014}. A striking discovery has been
that, 
under appropriate forcing, quasi-classical behavior arises
displaying statistical properties characteristic of
ordinary turbulence; an example is the celebrated
Kolmogorov $-5/3$ scaling of the energy spectrum 
\cite{Barenghi2014a} which suggests the existence of 
a classical energy cascade from large to small length scales.
Under other conditions, a different kind of turbulence (called
`ultra-quantum turbulence' or `Vinen turbulence') 
has also been found \cite{Walmsley2008, Baggaley2012}, 
characterized by random tangles of vortices without large-scale,
energy-containing flow structures. Quantum turbulence experiments are also
performed in atomic Bose-Einstein condensates \cite{Henn2009c, Henn2009, Neely2013}; the relative
small size of these condensates
(compared to flows of liquid helium or of ordinary fluids)
limits the study of scaling laws 
but offers opportunities to
study minimal processes that also take place in larger systems
(e.g. vortex interactions, vortex reconnections, vortex clustering) 
with greater experimental 
controllability and more direct visualization than in liquid helium. 

Atomic condensates
are also ideal systems to study two-dimensional (2D) turbulence \cite{White2014}, a problem
with important applications to oceans, planetary atmospheres and astrophysics.
In these systems, reduced dimensionality may arise from 
strong anisotropy,  stratification or rotation
(via the Taylor-Proudman theorem). From the physicist's point of view,
the dynamics of 2D turbulence is very different from 3D \cite{Kraichnan1980}.
The existence (besides the kinetic energy) 
of a second inviscid quadratic invariant 
- the enstrophy - implies that
a downscale enstrophy transfer is accompanied by an upscale energy
cascade; in other words, in 2D turbulent flows the energy flows from
small to large length scales rather than 
vice-versa as in 3D turbulence.
With the possible exception of soap films \cite{Rivera1998},  2D flows which
can be created in the laboratory are only approximations. However, using
suitable trapping potentials, atomic condensates can be easily 
shaped so that vortex dynamics is 2D rather than 3D.
Unlike liquid helium, in atomic condensates
2D quantum vortices can be directly imaged,
and, unlike classical systems, the motion of such 2D vortices
is not hindered by viscous effects or friction with the substrate.

Several works have explored the generation of turbulence in 2D condensates.
The 2D energy spectrum and scaling laws have been computed in numerical simulations 
\cite{Numasato2010, Parker2005, Nowak2011a}, and the problem of what should be the quantum analogue of 
the classical enstrophy has been raised.
In Ref.~\cite{Neely2013} vortices 
were nucleated by small-scale stirring of a laser spoon, after which a 
persistent current was verified both experimentally and through 
numerical simulations, suggesting transfer of incompressible 
kinetic energy from small to large length scales. Emergence of large-scale
order from vortex turbulence was also observed \cite{Simula2014} as 
predicted by the `vortex gas' theory of Onsager. 
A similar set-up was used to explore vortex shedding and annihilation 
processes in both experiments \cite{Kwon2014,Kwon2015} and 
simulations \cite{Stagg2015}. The effect of stirring laser beams with 
different shapes or along different paths
was investigated in Refs.~\cite{White2012, Reeves2012, White2014a, Neely2010}. 
However, in all of cited cases, vortices have always been 
generated in such a way that the number of positive and negative vortices
is approximately the same; in other words, all vortex configurations
which have been investigated had approximately zero polarization.
Since irrotational flow is a hallmark property of superfluidity,
the polarization of the vortex configuration (i.e. the relative proportion
of positive and negative vortices) plays the role of 
net rotational angular velocity $\Omega$ of a classical fluid, 
so it is important to explore its effects on the properties of turbulence.

In this work, we propose a new scheme for generating 2D quantum
turbulence in atomic condensates.
The novelty of our scheme, which is based on a giant vortex
as the initial state, is control
over polarization of the turbulence, which can be interpreted as
the classical rotation of the entire flow.  One of the most important
properties of turbulence is its decay, because the growth of the turbulence or
its character in a steady state may depend on how it is forced, whereas the
decay is an intrinsic property of the dynamics. We shall report the decay
of 2D quantum turbulence as a function of the polarization.

\section{Giant vortex and small pins}

Multicharged vortices with circulations as 
large as 60 quanta have already been produced in condensates
using dynamical methods, 
as consequences of rapid rotations of the confining trap \cite{Engels2003}. 
Another route to achieve these highly excited states is 
using phase-engineering techniques, such as those described 
in Refs~\cite{Leanhardt2002,Nakahara2000,Isoshima2007,Mottonen2002}. 
In these cases, quanta of angular momentum are added to the condensate
by adiabatically inverting the direction of the magnetic bias-field 
which composes the usual Ioffe-Pritchard magnetic trap. Up to this 
date, only charges below 10 quanta were produced using their 
proposed set-ups. However, an improvement on the method, known as the
`vortex-pump', has been described
in Refs.~\cite{Mottonen2007,Kuopanportti2010e,Kuopanportti2013}. 
In practical terms, a hexapole magnetic field is superposed to 
the Ioffe-Pritchard magnetic trap, allowing vorticity to be 
cyclically pumped into the condensate, and generating giant vortices. 
Progress in this direction has been done in recent 
experiments with synthetic magnetic monopoles  \cite{Ray2014}. 

A giant vortex at the center of a harmonically trapped condensate
can be described by a single-particle 
wave-function of the form $\psi(\textbf{r})=f(r)e^{i\kappa\phi}$, where 
$f(r)$  is the wave-function's amplitude,
${\bf r}=(r,\phi,z)$ is the position in cylindrical coordinates,
and a large winding number $\kappa$ corresponds to a large angular momentum.
Such giant vortices are dynamically unstable 
\cite{Kuopanportti2010e, Kuopanportti2010b}, and split
into singly-quantized vortices.  Being parallel to one another, these 
singly-quantized vortices impose a strongly azimuthal flow
to the condensate. During the following evolution,
some vortices of the opposite polarity may be generated
by occasional large-amplitude density waves, but
these events are rare, and do not change the main property of the flow
resulting from the decay of a giant vortex
configuration: the strong polarization of the vorticity - 
almost all vortices have the same sign.

The scheme that we propose uses blue-detuned 
lasers \cite{Kwon2015} to perturb this initial state with two diametrically 
opposite laser beams, creating thin obstacles (which we refer as pins) with width $\sigma$ of the order
of magnitude of the healing length $\xi$ (two pins
are enough to homogenize the vortex distribution). 
The pins perturb the initial giant vortex, accelerating its decay;
they also deflect the large azimuthal flow, generating
vortex-antivortex pairs \cite{Frisch1992, Winiecki1999,Berloff2000,Stagg2015}.
To control the effect of the pins, we move them
at constant angular velocity $\omega$ in the direction
opposite to the main azimuthal flow. 

\section{Model}

The dynamics of our system is dictated by the 
2D Gross-Pitaevskii equation (GPE). We introduce
dimensionless variables based on the trapping potential of frequency
$\omega_0$, measuring
times, distances, and energies  
in units of $\omega_0^{-1}$, $\sqrt{\hbar/m\omega_0}$ 
and $\hbar\omega_0$ respectively, 
where $m$ is the mass of one atom and $\hbar$ is the reduced 
Planck's constant. The resulting dimensionless GPE is

\begin{equation}
\label{gp}
i\frac{\partial\psi}{\partial t}=
\left(-\frac{1}{2}\nabla^{2}+V
+C\left|\psi\right|^{2}-\mu\right)\psi,
\end{equation}

\noindent
where the time-dependent wavefunction $\psi({\bf r},t)$ is 
normalized so that $\int \vert \psi \vert^2 d^3r=1$.
The external potential is
$V({\bf r},t) = V_\mathrm{trap}({\bf r}) + V_\mathrm{pins}({\bf r}, t)$, 
where $V_\mathrm{trap}({\bf r})=(x^2+y^2)/2$ and 
$V_\mathrm{pins}({\bf r}, t) = V_{+}(\mathbf{r}, t) + V_{-}(\mathbf{r}, t)$ 
represent respectively the trapping potential which confines the
condensates and the pins which perturb the initial giant vortex. 
The terms 
$V_{\pm}(\mathbf{r},t)= 
V_0\exp\left\{-\vert {\bf r}-{\bf r}_{\pm}(t)\vert^2/2\sigma^2\right\}$ 
with ${\bf r}_{\pm}(t)=\left[\pm x_0 \cos{(\omega t)},y_0\sin{(\omega t)}\right]$
are diametrically-opposite, thin,
Gaussian potentials of width $\sigma=\xi$
which rotate clockwise
(against the flow of the initially imposed giant vortex) 
at constant angular velocity $\omega$. 
The quantity $C = 2\sqrt{2\pi}N(a/a_z)$ parametrizes the two-body
collisions between the atoms, where $N$ is the total number of atoms, $a$ the scattering length, and $a_z$ the axial harmonic oscillator's length; we choose $C=17300$.
The chemical potential $\mu$ is introduced to guarantee 
normalization of the wave-function, and the amplitude of the pins is
$V_0 \approx 1.43 \mu$. In homogeneous systems ($V=0$) the healing length 
is found by balancing kinetic and interaction energies terms in the GPE.
In a harmonically trapped condensate, the healing length 
can be defined with reference to the density at the center of the trapped
condensate in the absence of any vortex or hole. 
In our dimensionless units,
we obtain $\xi\approx 0.13$,
 and $r_{TF}\approx 74\xi$ for the Thomas-Fermi radius.

Our choice of dimensionless parameters corresponds to typical
\cite{Kwon2014,Kwon2015} experiments with  $^{23}$Na condensates
(scattering length $a=2.75~{\rm nm}$,
atom mass $m = 3.82 \times 10^{-26} {\rm kg}$) with 
$N=1.3 \times 10^6$ atoms,
radial and axial trapping frequencies
$\omega_0=2 \pi \times 9$ Hz and $\omega_z = 2\pi\times 400$ Hz,
radial and axial harmonic oscillator's lengths 
$a_0 = \sqrt{\hbar/m\omega_0}\approx 7.1 ~\mu{\rm m}$ and
$a_z=\sqrt{\hbar/m\omega_z}\approx 1.0$ $\mu$m, for which the dimensional
healing length is $\xi = 0.13 a_0 \approx 0.9$ $\mu$m; the laser beam would have then a Gaussian $1/e^2$ radius of $w_0 = 2\sigma \approx 1.8$ $\mu$m. Blue-detuned Gaussian laser beams have been used as pins in a series of experiments with highly-oblate BECs \cite{Kwon2015,Neely2010,Desbuquois2012}. Particularly in \cite{Desbuquois2012}, a laser beam of width $w_0 \approx 2$ $\mu$m was used to stir a 2D $^{87}\mathrm{Rb}$ condensate, similarly to what we propose, maintaining a circular motion with the help of piezo-driven mirrors.

In order to define our initial state, a circulation of $37$ quanta 
(i.e. winding number $\kappa=37$) is initially
imprinted around the center of the
Thomas-Fermi profile, thus imposing an initial counter-clockwise circular 
flow.  Changing $t$ into $-it$ in Eq.~(\ref{gp}), we shortly evolve the state for $t=0.09$ in  
imaginary-time description, guaranteeing a fixed phase of $2\pi\kappa$ 
in the center of the condensate and adjusting the density to the presence 
of the pins. We then compute the evolution in real time.

By substituting $t \rightarrow (1-i\gamma)t$ in Eq.~\ref{gp}, we are left with a phenomenological dissipative GPE (dGPE), where $\gamma$ is a dissipation constant which models the interaction of the condensate 
with the surrounding thermal cloud. This equation can be used to investigate the effect of finite temperature in our system. With this aim, we also repeat our simulations using dGPE with $\gamma = 3.0\times 10^{-4}$, a typical value of dissipative parameter \cite{Choi1998, Tsubota2002, Bradley2012}, particularly chosen for the for experimental realistic case found in current experiments \cite{Kwon2014,Stagg2015}. Summarizing beforehand, we find the same overall behavior for both dissipation-less and this specific dissipative case.

All numeric simulations are performed in the 2D domain
$-25\leq x,y \leq 25$ on a $512 \times 512$ grid using 
the 4th order Runge-Kutta 
method in Fourier space with the help of XMDS \cite{Dennis2013}.  

\section{Results}

\subsection{Creating Polarized Flow}
\begin{figure}[h!]
  \centering
  \subfigure{\includegraphics[width=0.16\textwidth]{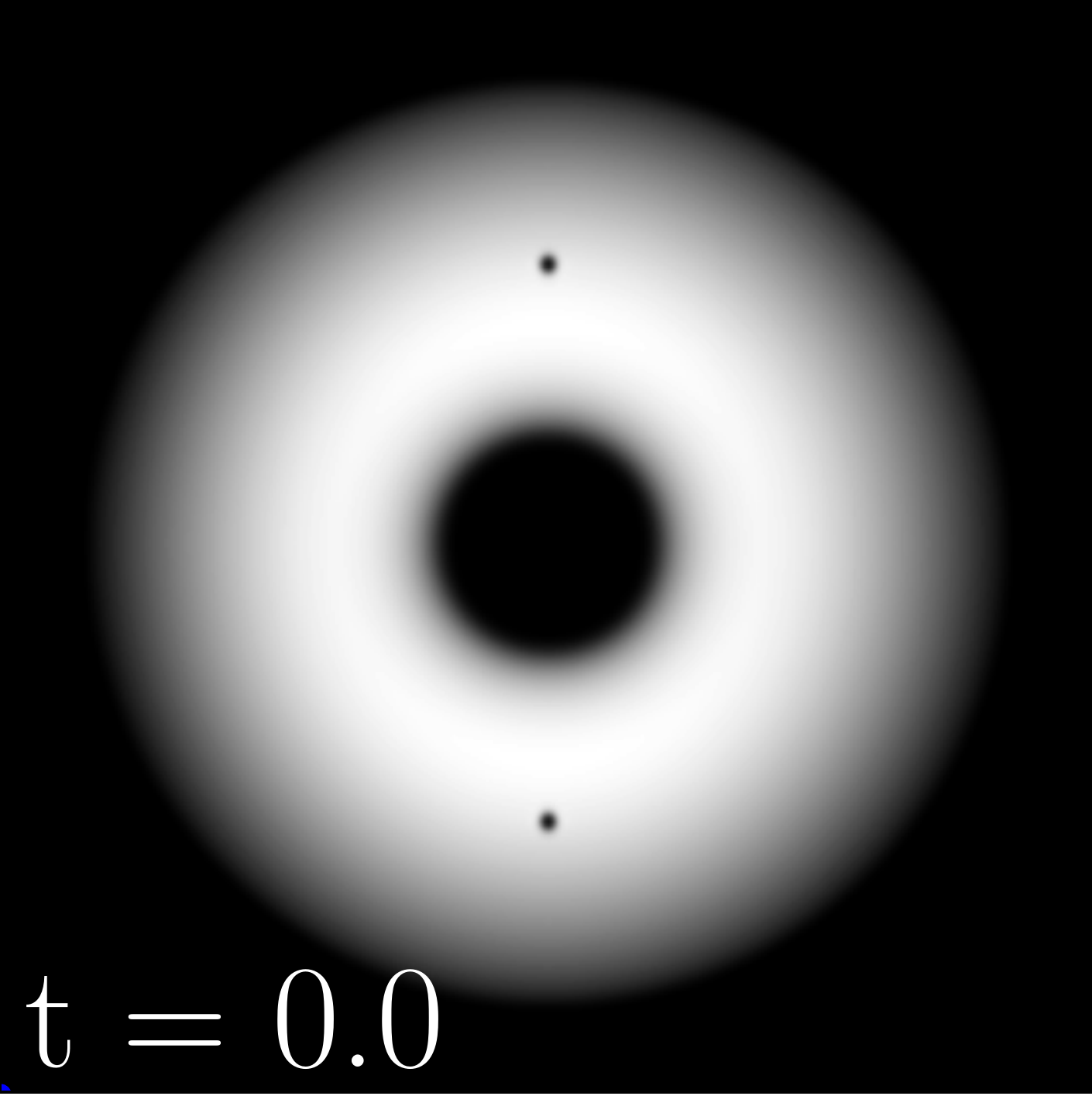}}
  \hspace{-0.33cm}
  \subfigure{\includegraphics[width=0.16\textwidth]{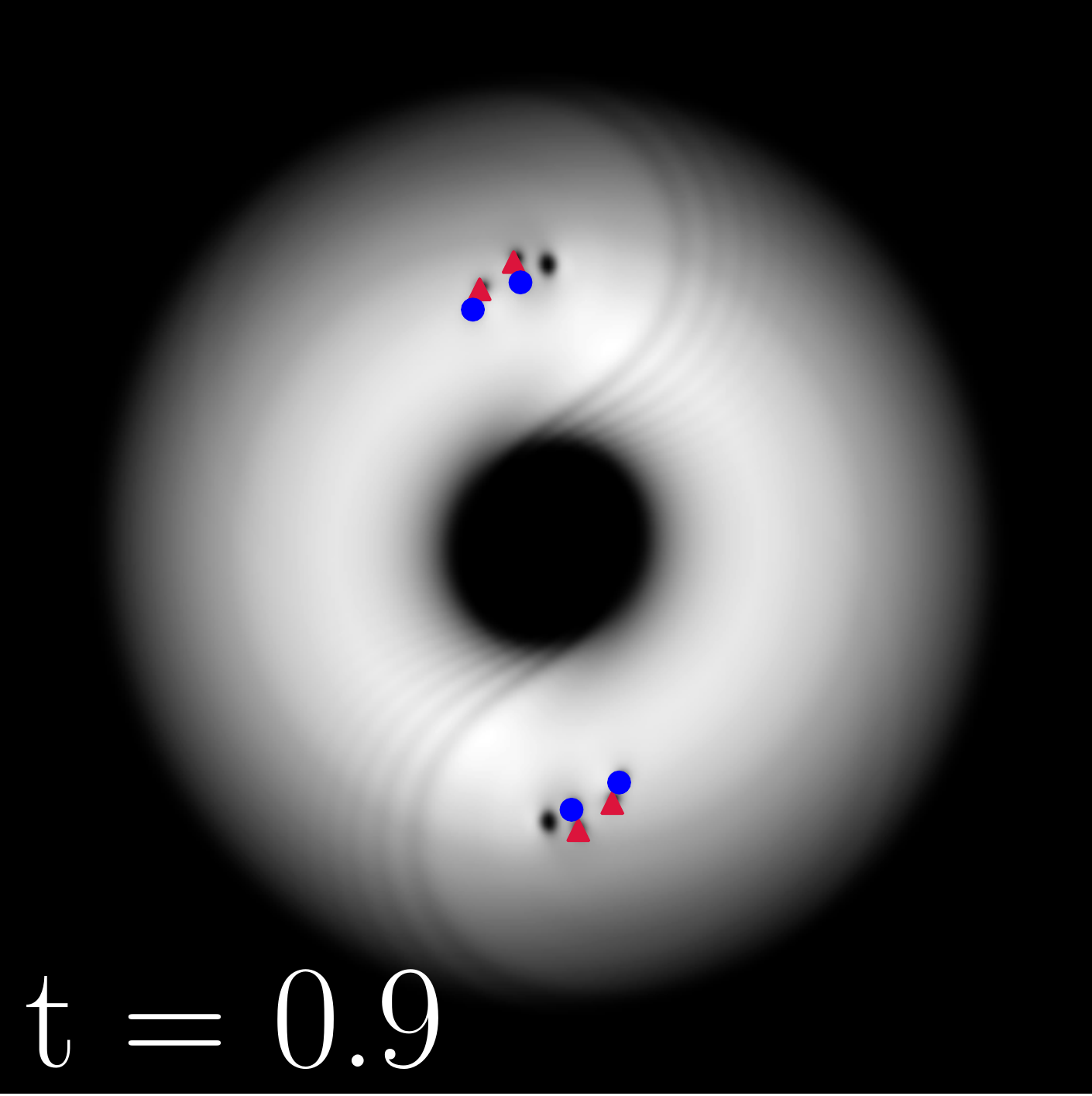}}
  \hspace{-0.33cm}
  \subfigure{\includegraphics[width=0.16\textwidth]{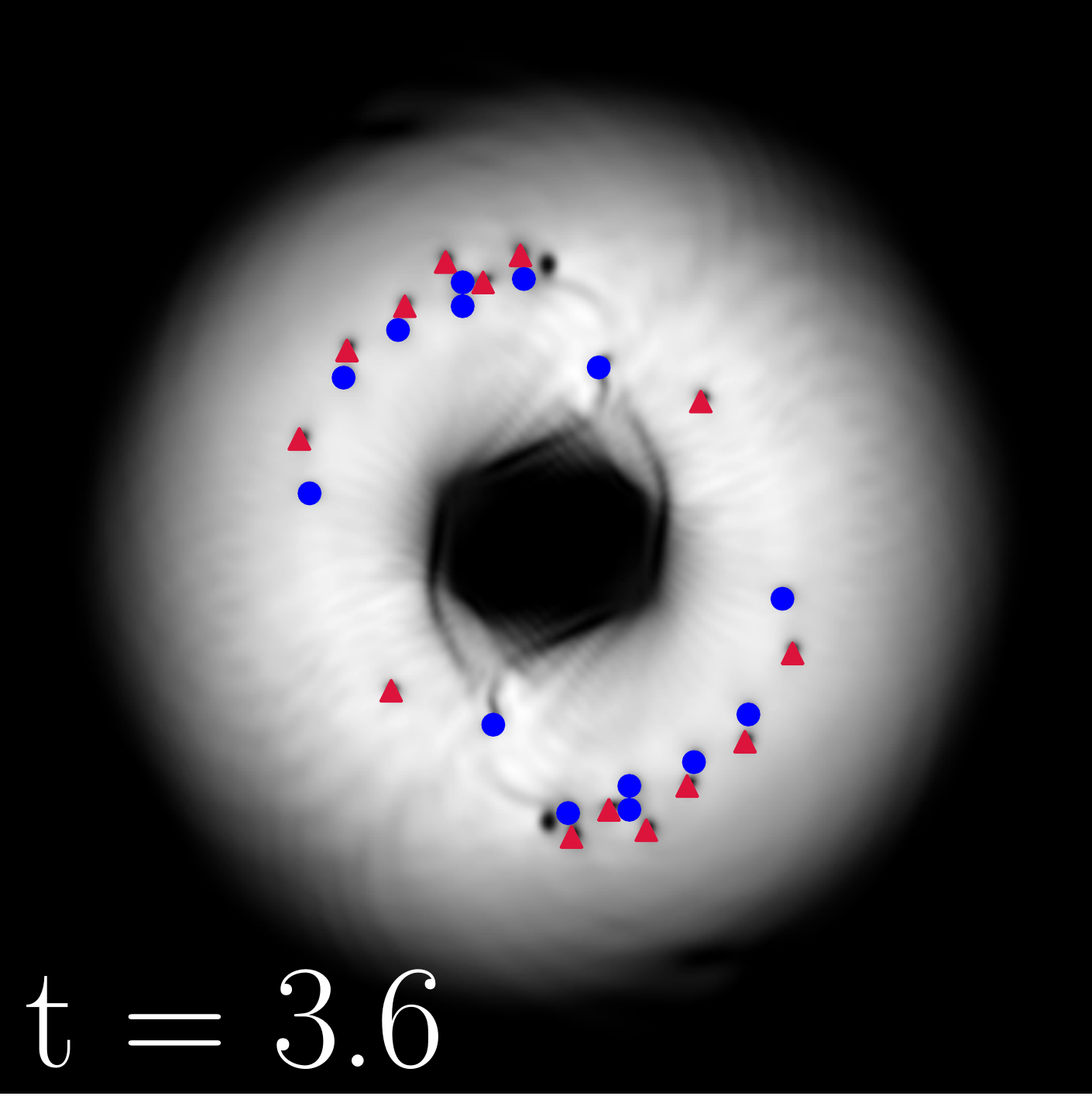}}
  \par\vspace{-0.38cm}
  \subfigure{\includegraphics[width=0.16\textwidth]{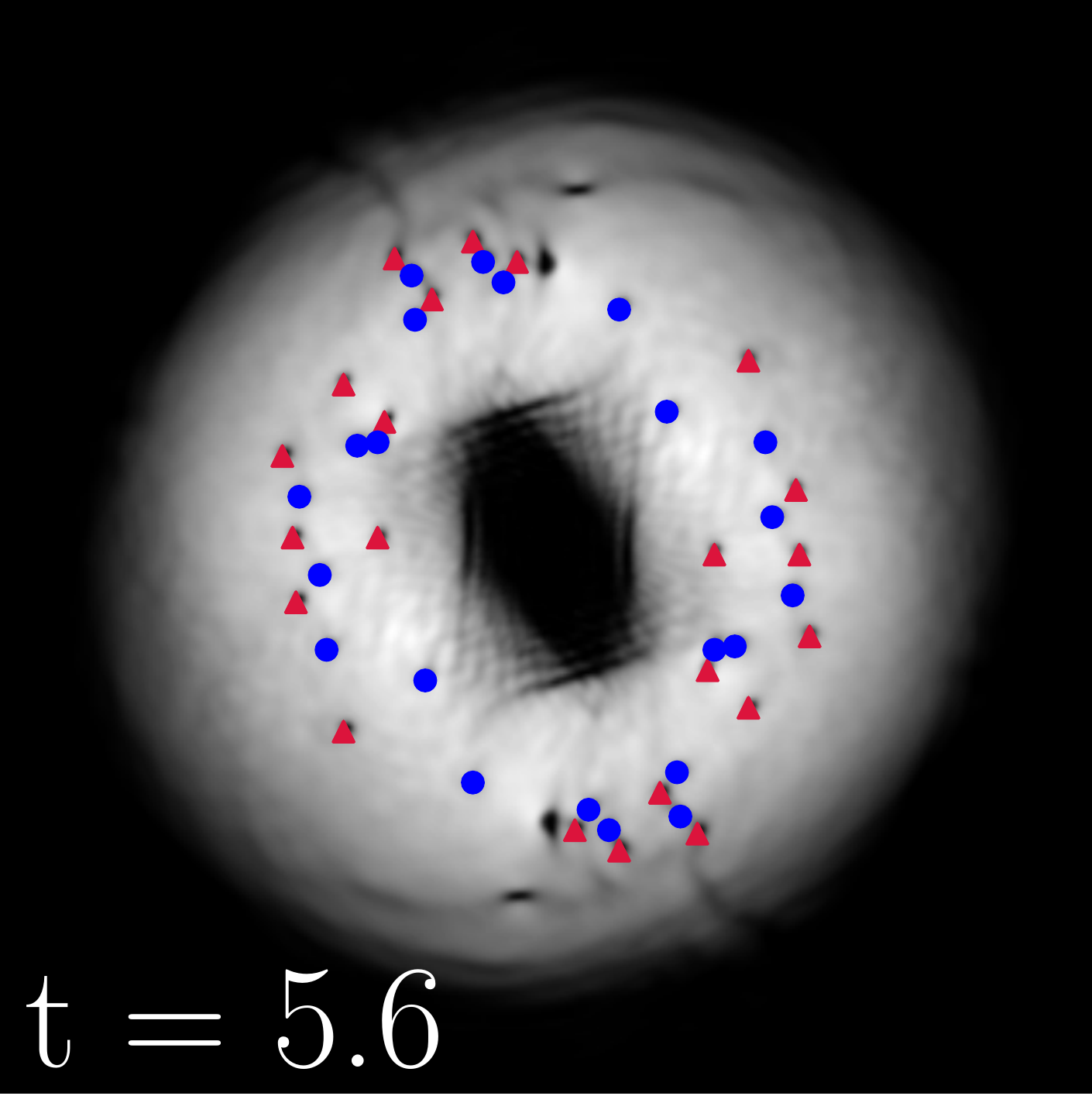}}
  \hspace{-0.33cm}  
  \subfigure{\includegraphics[width=0.16\textwidth]{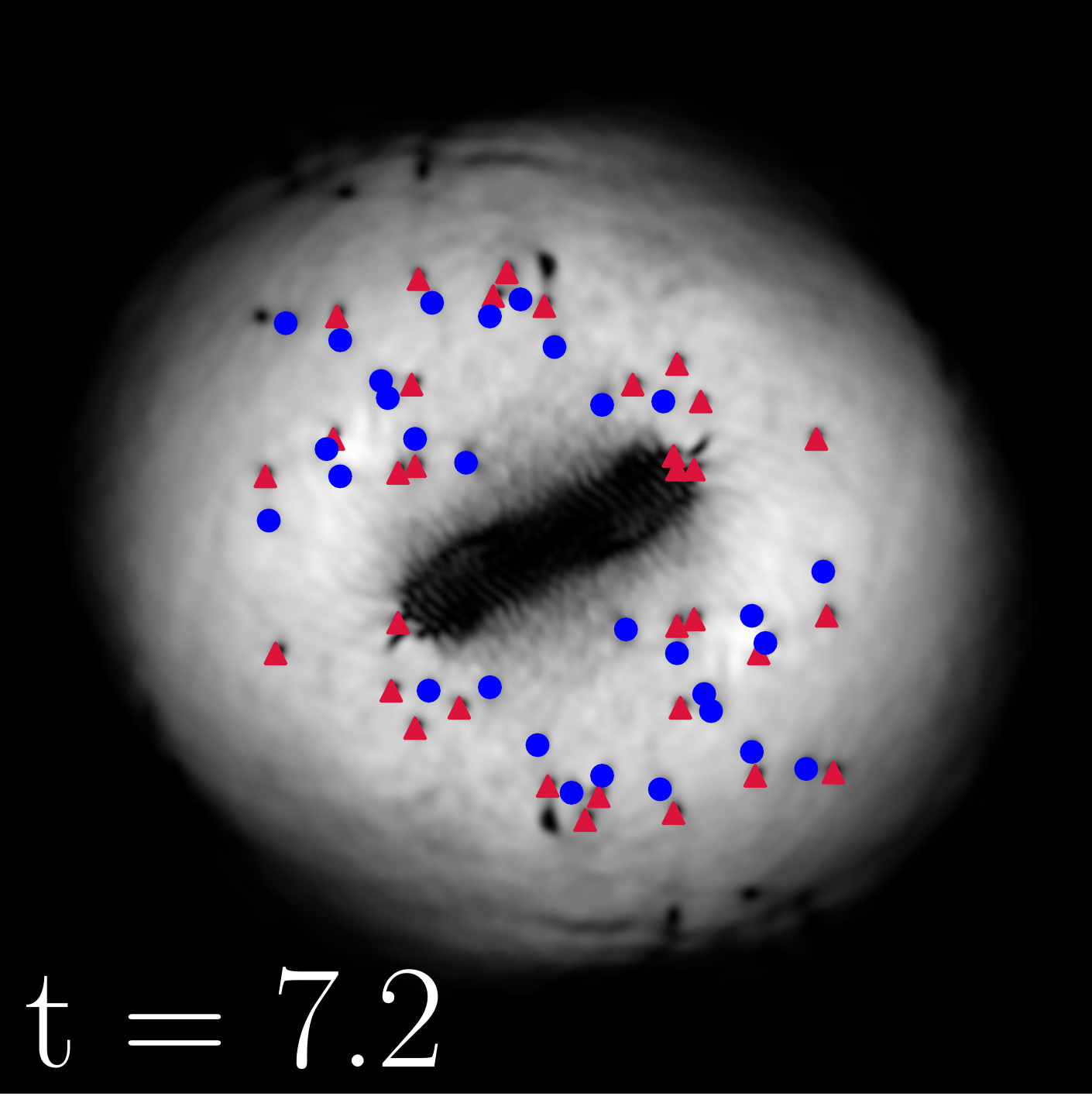}}
  \hspace{-0.33cm} 
  \subfigure{\includegraphics[width=0.16\textwidth]{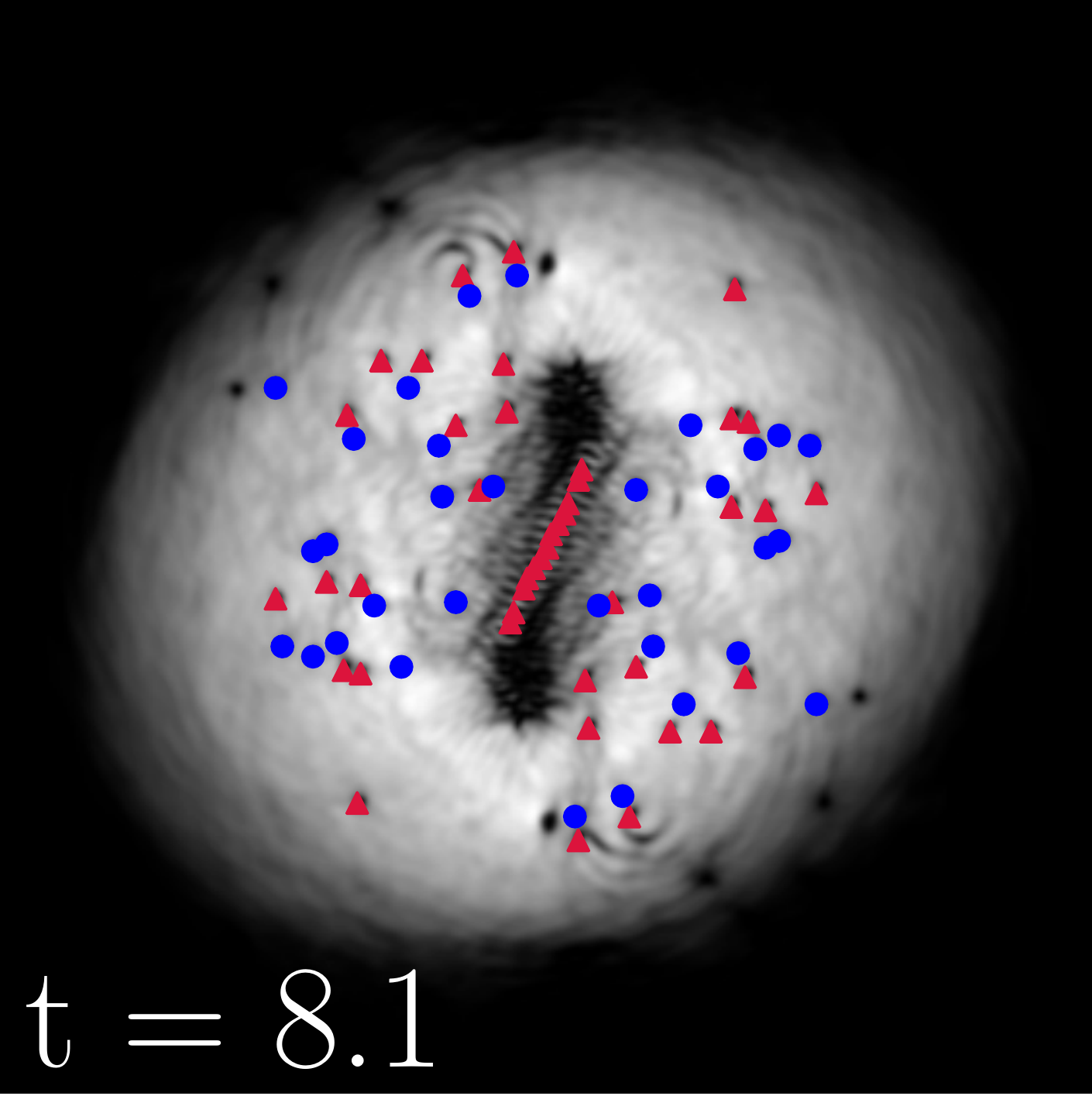}}
  \par\vspace{-0.38cm}
  \subfigure{\includegraphics[width=0.16\textwidth]{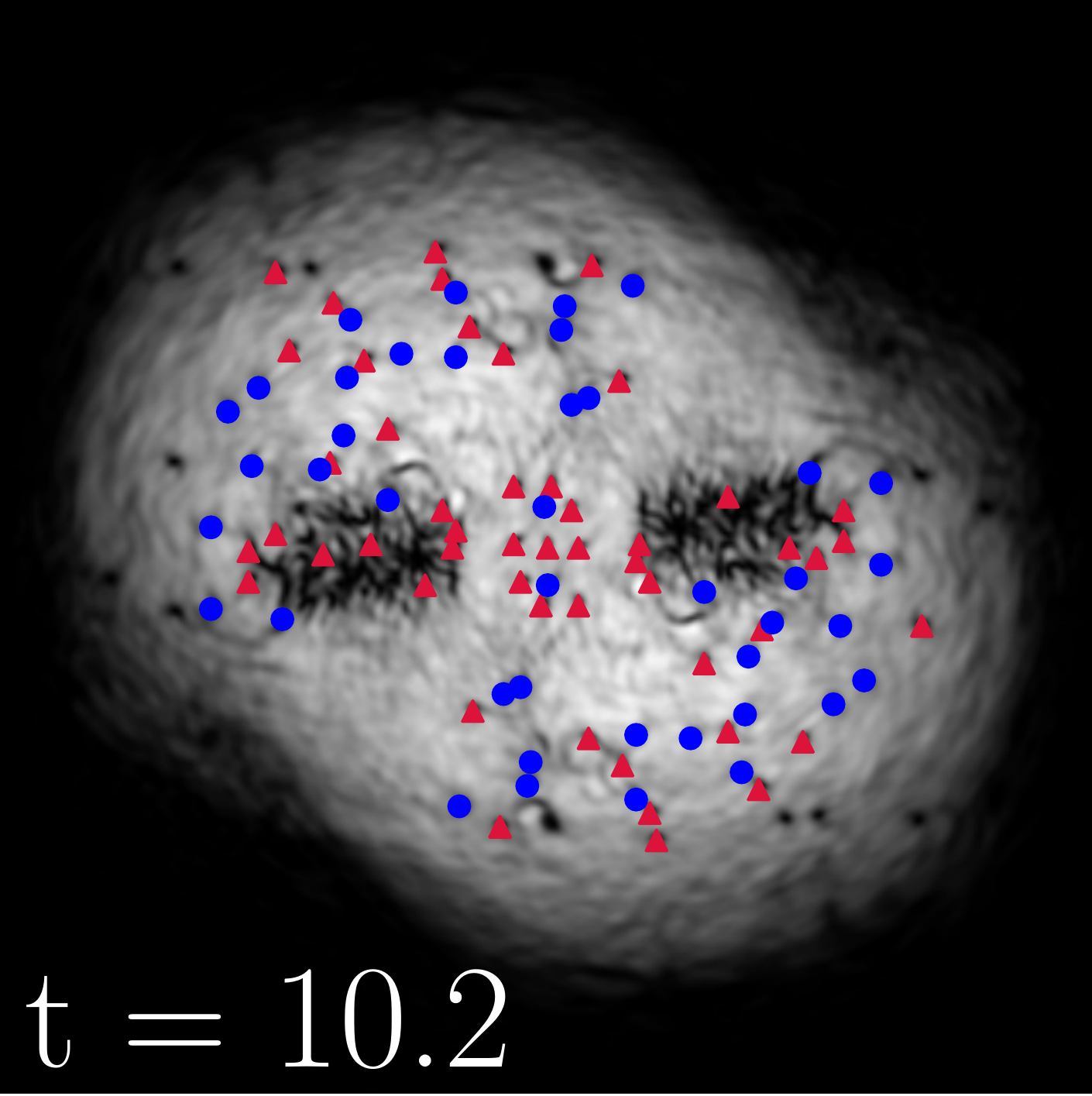}}
  \hspace{-0.33cm} 
  \subfigure{\includegraphics[width=0.16\textwidth]{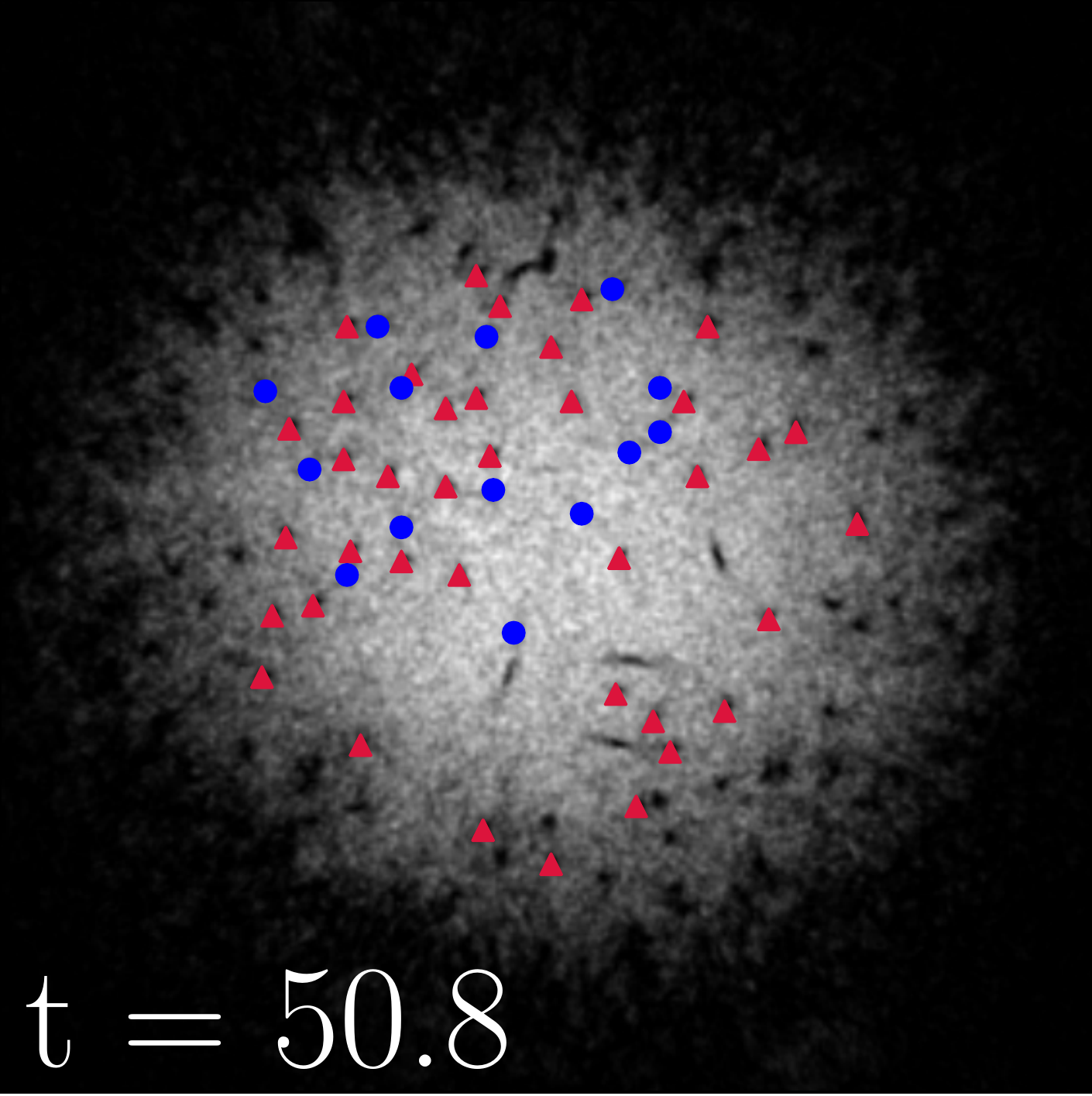}}
  \hspace{-0.33cm}   
  \subfigure{\includegraphics[width=0.16\textwidth]{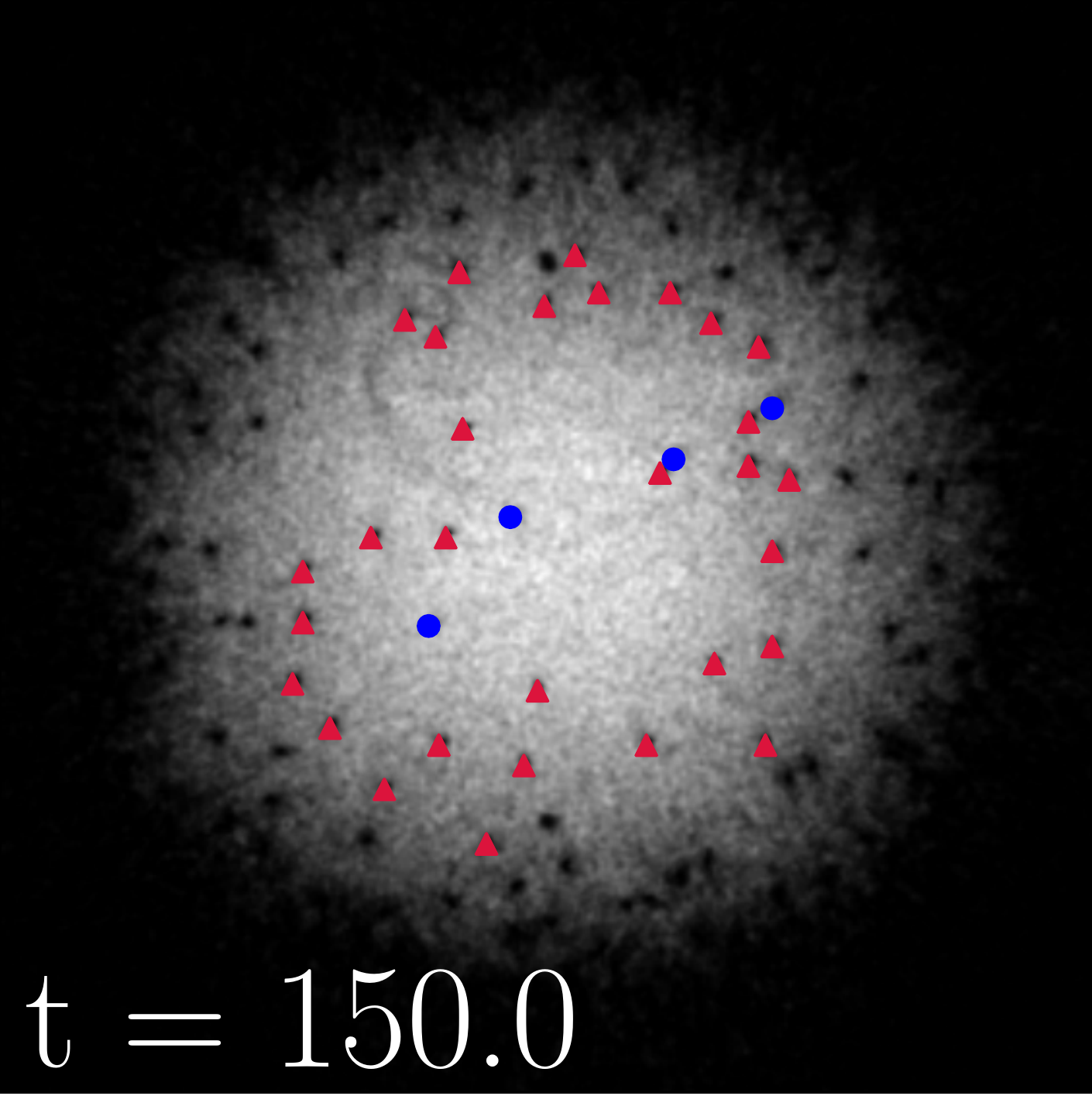}}%
\caption{
(Color online). 
Density plots of the condensate at different
times $t$ for $\omega=0$ (non-rotating obstacles). Regions of large/low density
are displayed in white/black respectively.
Red triangles and blue circles identify  positive-charged and negative-charged
vortices respectively. The giant vortex decays by injecting a large number 
of singly-quantized (positive) vortices into the condensate,
whilst the pins generate vortex pairs, as can be clearly seen at time 
$t = 0.9$. 
}
\label{fig1} 
\end{figure}

We simulate the real-time evolution of the system for different 
values of the pins' angular velocity: 
$\omega = 0$, $\pi/16$, $\pi/8$, $\pi/6$, and $\pi/4$. 
A series of snapshots for the case of $\omega = 0$ is shown in 
FIG.~\ref{fig1} to exemplify a typical run. The initial large hole
at the center of the figure is the core of the giant vortex.
The two small holes (north and south of the giant hole) are the two stationary
pins.
The critical velocity $v_c$ for
the creation of a vortex-antivortex pair
depends on the barrier's shape \cite{Pinsker2014,Stagg2015} and also on 
inhomogeneities of the system \cite{Kwon2015}. 
Typically $v_c/c \sim 0.1 - 0.4 $ for infinitely high cylindrical barriers, 
where $c$ is the local speed of sound. Since our barriers (the pins) 
are either stationary or
rotate against the main flow, vortex shedding is a dissipative 
mechanism which slows down the superfluid's azimuthal flow and removes
angular momentum.

Besides generating vortices of opposite sign, the pins act as
perturbation to the giant 
vortex and quicken its decay process; for example, a 
wave front which perturbs the core of the giant vortex is visible at time
$t=0.9$ in FIG.~\ref{fig1}. The decay of the giant vortex takes place
via deformation of the core, which becomes elliptical before vanishing,
and injecting 
a large number of positive, singly-quantized vortices into the condensate.
At the same time, vortex-antivortex pairs are created by the flow past
the pins. This process continues until the large azimuthal flow is 
lower than the critical velocity $v_c$; at that point the giant vortex has disappeared, and the pins are practically unable to
generate further vortices. Therefore, after this slowdown and due to their small sizes, the pins are practically irrelevant to the vortex dynamics (apart from occasional creation of pairs in the fast rotating case, $\omega = \pi/4$). In spite of that, in order to study the vortex number decay, we simply remove them at $t=82$ and allow for longer simulations.

\begin{figure}[ht!]
\begin{center}
\includegraphics[width=0.5\textwidth]{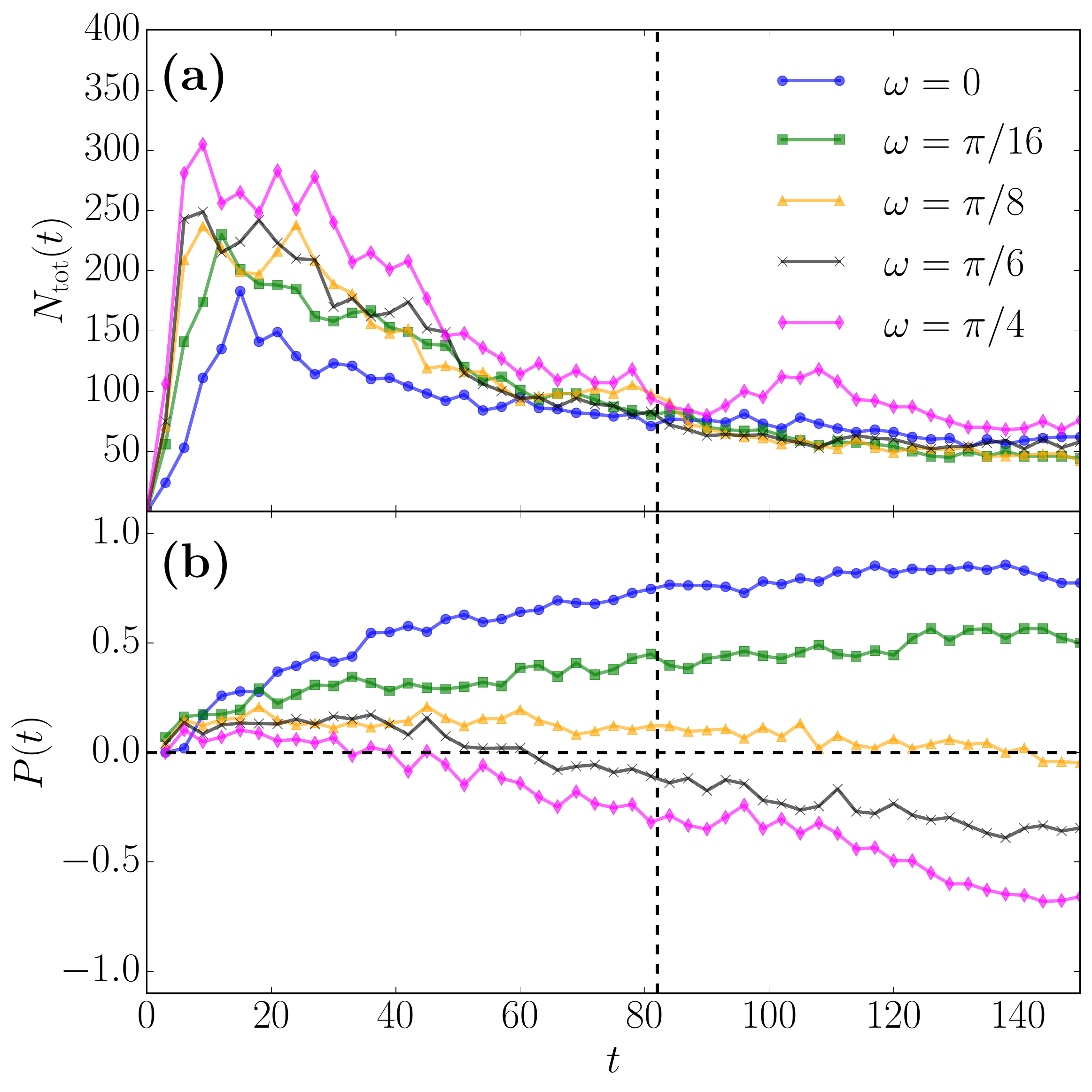} 
\caption{
(Color online).
\textbf{(a)} total number of vortices, $N_\text{tot}$ vs time $t$;
\textbf{(b)} polarization $P$ vs time $t$. The vertical dashed line
marks the time ($t=82$) we use to make initial states for longer simulation without the pins. The curves are distributed in increasing value of $\omega$ from bottom to top, for the top plot, and conversely, for the bottom plot. In (a), at $t\approx 8$, from bottom to top, the curves refer respectively to $\omega=0,\pi/16,\pi/8,\pi/6, \text{and }\pi/4$. In (b), from top to bottom, the curves refer respectively to the same latter series of increasing values of $\omega$. 
}
\label{fig2}
\end{center}
\end{figure}

We perform a phase-unwrapping procedure and, by detecting windings of $\pm 2\pi$ around small closed paths (plaquettes) on the phase-profile \cite{White2012}, we are allowed to count the
numbers $N^+$ and $N^-$ of positive and negative
singly-quantized vortices in the system (anticlockwise and clockwise
circulation respectively). This vortex detection algorithm uses a density-cut criterion ($\sim 0.75$ of $\psi$'s mean-value) to avoid detection of ghost-vortices. Given the initial giant vortex (which is multicharged and therefore not detected by our vortex-detecting algorithm),
depending on the value of $\omega$, there can be an imbalance of $N^{+}$ 
and $N^{-}$ throughout the evolution. Vortices can be expelled from the 
condensate due to their mutual interaction, spiral out of the
condensate because of 
dissipation, or undergo vortex-pair annihilation processes. 
In our particular finite-temperature simulation, we verify that the chosen experimental realistic value of the dissipation parameter $\gamma$ is small enough
that, on the time scale analyzed (and compared to the dissipation-less 
simulations),
dissipation-induced spiraling out of individual vortices is less 
important than vortex interactions or annihilations.

\begin{figure}[ht!]
\begin{center}
\includegraphics[width=0.5\textwidth]{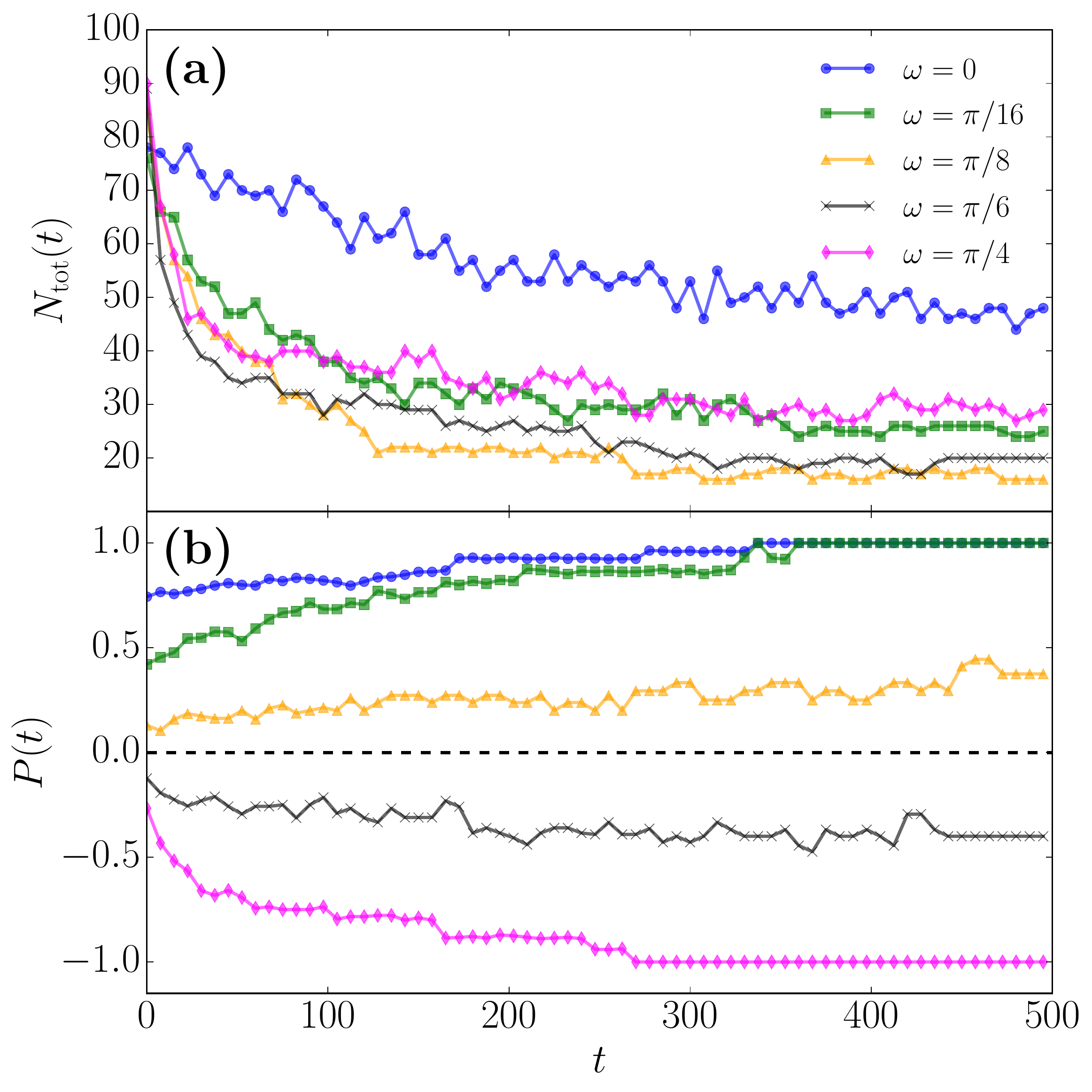} 
\caption{
(Color online).
\textbf{(a)} total number of vortices $N_\text{tot}$; \textbf{(b)} polarization $P$ vs time $t$, from initial states created at $t=82$ in the previous stirring process (FIG.~\ref{fig2}), labeled by the angular velocities which generated them.
The pins are removed and we evolve those states longer in time to study the vortex number decay. In (a), at $t\approx 150$, from bottom to top, the curves refer respectively to $\omega=0,\pi/4,\pi/16,\pi/6, \text{and }\pi/8$. In (b), from top to bottom, the curves are labeled as in FIG. 2 (b). 
}
\label{fig3}
\end{center}
\end{figure}

After the decay of the initial giant vortex, 
the imbalance of positive and negative vortices 
is measured by the polarization  
$P = (N^{+}-N^{-})/(N^{+}+N^{-})$, which takes
maximum/minimum values ($P=\pm 1$) if all vortices have positive/negative
sign. 
FIG.~\ref{fig2} shows the time evolution of the total 
number of vortices $N_\text{tot}(t)=N^{+}+N^{-}$ and of the
polarization $P(t)$ under influence of the obstacles (present throughout the whole evolution) with angular 
velocity
$\omega$. The top part (a) of the figure  shows that $N_\text{tot}(t)$ increases
with $\omega$. It is apparent that, by choosing $\omega$, we can
control the polarization. We use this tunable mechanism to create initial vortex distribution (without the pins) as shown in FIG.~\ref{fig3}, which plots
$N_\text{tot}(t)$ and $P(t)$ for  initial states taken from instant $t=82$ of FIG.~\ref{fig2}. Clearly,
by tuning $\omega$ we can produce a condensate free of external holes (the
giant vortex or the obstacles) with approximately the desired vortex
polarization.

By numerically detecting each vortex and its trajectory, we determine
$N_\text{tot}$ at each time step. By subtraction from the initial total vortex number, $N_0 = N_\text{tot}(0)$, we can infer the number of vortices which have drifted out of the
condensate, $N_\text{dr}$, and the number of vortices which have
disappeared in annihilation events, $N_\text{an}$,
colliding with vortices of opposite sign. We find that such 
vortex-antivortex annihilation events generate density waves, as already
reported \cite{Leadbeater2001,Parker2004,Stagg2015}, turning kinetic
energy into sound energy.  The reverse mechanism is also possible
\cite{Berloff2004} and in our 2D case takes the form of
vortex-antivortex creation events, which we observe. Creation events
occur when
the motion of the vortices induces a sufficiently deep density
wave, or when a large amplitude wave approaches the edge of the condensate
where the local speed of sound $c$ is less than in the central region.
We have also observed annihilations events immediately followed by
creation events: this sequence happens when a vortex collides with
an antivortex, producing 
a large sound wave, which almost immediately generates a new vortex-
antivortex pair, due to the changing value of the local ratio $v_c/c$;
this effect happens near the condensate's edge.

\begin{figure*}[ht]
  \includegraphics[width=\textwidth]{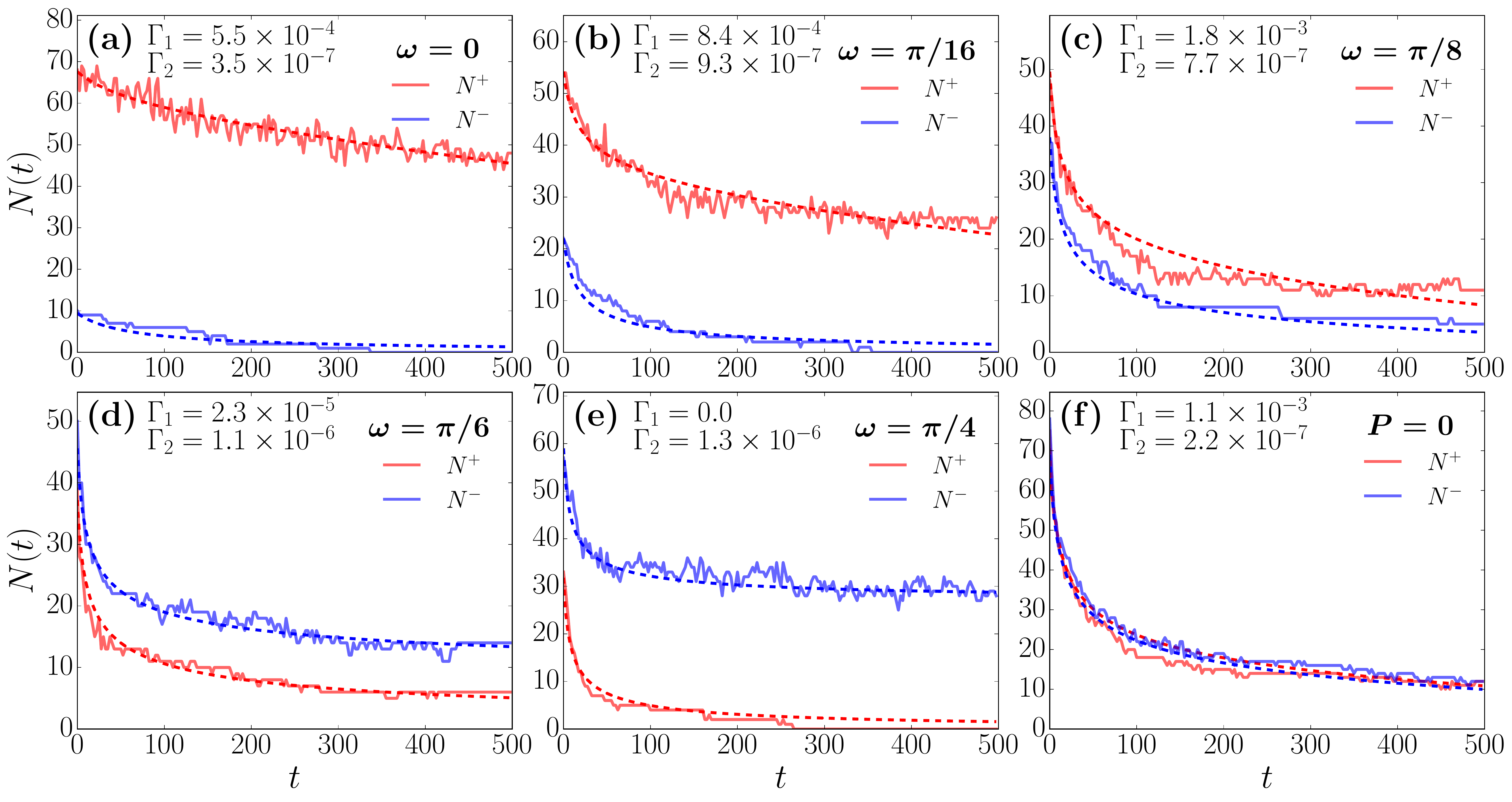}
\caption{
(Color online) Positive and negative vortex number $N^+$ and $N^-$ decay as a function of time $t$ for cases (a) $\omega = 0$;  (b) $\omega = \pi/16$; (c) $\omega = \pi/8$; (d) $\omega = \pi/6$; (e) $\omega = \pi/4$; and (f) phase-imprinted $P=0$. The dashed lines are the respective fits for the numerical data (full lines) with the fitting parameters $\Gamma_1$ and $\Gamma_2$ appearing at top part of each plot.}
\label{fig4} 
\end{figure*}

\subsection{Polarized Turbulence Decay}
Starting from  $t>82$ (when we remove the pins and start a new simulation) we can examine
whether there is a simple
law for turbulence decay in 2D condensates. 
It has been suggested \cite{Kwon2014,Stagg2015} that the decay rate of the 
total number of vortices is not exponential and can be
phenomenologically described by the logistic equation 
\begin{equation}
\frac{dN_\text{tot}}{dt}=
-\Gamma_1 N_\text{tot} - \Gamma_2 N_\text{tot}^2,
\label{logistic}
\end{equation}

\noindent
where the linear term should refer to vortex drifting out of the
condensate, the non-linear term arises from
vortex-antivortex annihilation events,
and the coefficients $\Gamma_1$ and $\Gamma_2$ are rates to be determined. 
We find that the solution of the logistic equation fits our decays for
$t>82$ (after pins removal) fairly well. However, the fitting parameters for the
linear rate $\Gamma_1$ were shown to be negative in most cases, corresponding
to positive growth. 
After the pins are removed, there is no reason or evidence to expect 
a vorticity source term, apart from occasional creation of vortex-antivortex
pairs mentioned above. Therefore, a naive association of 
the linear term of the logistic equation
with vortex drifting out of the condensate
due to dissipation effects is not appropriate in our case.

Instead we propose a modified approach to the phenomenological description, focusing on changing the non-linear contribution to the rate equation. It is clear that we must incorporate polarization in the description. In Ref. \cite{Kwon2014}, the authors include the non-linear term based on the intuitive assumption that annihilation processes are likely to occur with a rough dependence on the number of dipole pairs (composed of a positive and a negative vortex) which can be formed. In a zero-polarized system this is of order $\propto  N^2$. Following a similar reasoning, for a polarized system, the non-linear term should then be $\propto N^{+}N^{-}$. This consideration takes into account the fact that fewer vortex-dipole pairs are formed if $P\neq 0$ and implicitly allows for polarization's time dependence.


With this motivation, we found that the phenomenological model based on the coupled differential equations
\begin{equation}
\begin{aligned}
\frac{dN^{+}}{dt}&=
-\Gamma_1 (N^{+})^{3/2} - \Gamma_2 (N^{+}N^{-})^2, \\
\frac{dN^{-}}{dt}&=
-\Gamma_1 (N^{-})^{3/2} - \Gamma_2 (N^{+}N^{-})^2, 
\end{aligned}
\label{decoupled}
\end{equation}
fitted well our simulation curves, giving non-negative values for both fitting parameters $\Gamma_1$ and $\Gamma_2$, as show the respective values inserted in the plots of FIG.~\ref{fig4}.

The $3/2$ scaling for the first terms on the right-hand side of Eqs.~\ref{decoupled} arises from a simple argument which assumes that: (1) a singly-charged vortex (with either negative or positive circulation, $\kappa = \pm 1$) moves radially with a typical velocity proportional to $v_r=1/(2\pi l)$, where $l=1/\sqrt{n}$ is the inter-vortex spacing and $n=N/(\pi R^2)$ is the vortex density in a condensate of radius $R$; (2) vortices are expelled from the cloud in a proportion $S$, which in first approximation is independent of the vortex number $N$ but may depend on the dissipative parameter $\gamma$. Therefore, in an infinitesimally thin annulus of area $dA = 2\pi R v_r dt$ close to the border of the condensate there is a loss of $dN = -S n dA$, and we get the scaling $dN/dt \propto N^{3/2}$.       

In order to account for the initial steep descent of some of the curves, it was necessary to square the polarization-dependent term of the equation, i.e. $N^{+}N^{-}$, otherwise the fits have shown to be very poor. Therefore, summing-up Eqs.~(\ref{decoupled}), the total number of vortices decays non-trivially as a function of polarization $P$ according to 
\begin{equation}
\frac{dN_\text{tot}}{dt}=
-\Gamma_1 N_\text{tot}^{3/2} - \Gamma_2f(t)N_\text{tot}^4,
\label{ntotdecay}
\end{equation}
where the time-dependent polarization appears in the function $f(t)\equiv (P(t)^2-1)^2/8$. The same scaling $N_\text{tot}^4$ for the vortex number decay rate was justified heuristically in the context of a quenched 2D homogeneous system in Ref.~\cite{Schole2012}.

From FIG.~\ref{fig4} we can identify cases (b) and (e) ($\omega=\pi/16$ and $\omega=\pi/4$),  (c) and (d) ($\omega=\pi/8$ and $\omega=\pi/6$), respectively as opposite-polarization counterparts; their decay curves are similar in behavior, while having a polarization mirror-symmetry. The main difference between these curves and their counterparts is the steeper number decay in initial times for (c) and (e). Calculating the number of vortices decayed due to annihilations, we found that this loss was always considerably less than the drifting. The latter is strongly induced by vortices interactions, which push each other out of the cloud. Since our chosen dissipation was shown to be ineffective in making vortices spiral-out of the condensate in the time-scale studied, we conclude that both linear and non-linear terms in Eq.~(\ref{ntotdecay}) have origins on vortex interactions. Case (e) ($\omega = \pi/4$) illustrates well this need for steeper than quadratic term in the rate equation, since it characterizes a purely non-linear decay (i.e. $\Gamma_1 =0$).

\begin{figure}[t!]
\begin{center}
\includegraphics[width=0.5\textwidth]{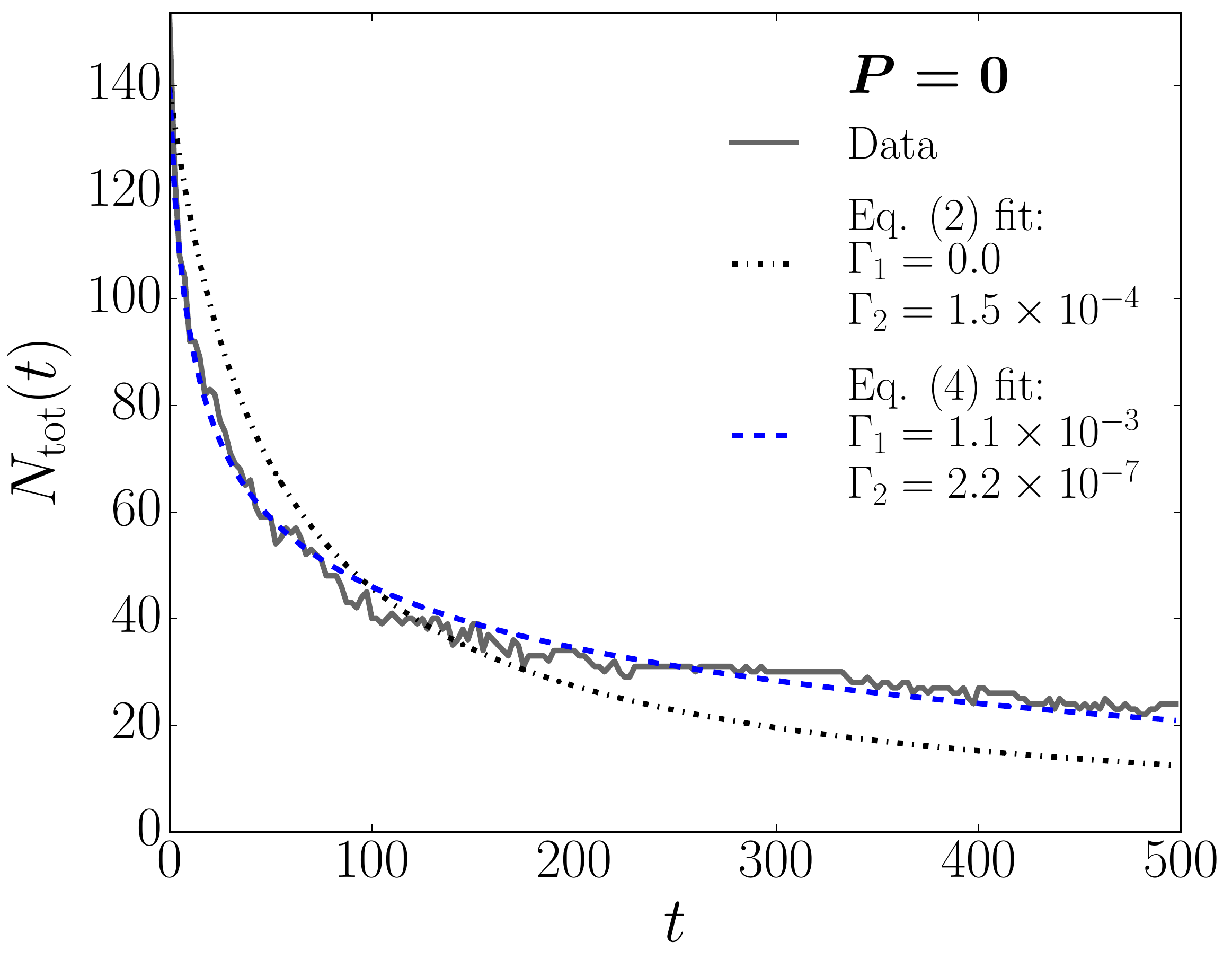} 
\caption{
(Color online) Total number of vortices, $N_\text{tot}$ vs time $t$ for the case where an unpolarized $P=0$ vortex distribution was created through phase-imprinting, in order to compare fits given by Eq.~(\ref{logistic}) (black dot-dashed line) and Eq.~(\ref{ntotdecay}) (blue-dashed line)
}
\label{fig5}
\end{center}
\end{figure}

Finally, in order to compare models from Eq.~(\ref{logistic}) with Eq.~(\ref{ntotdecay}) we performed numerical experiments in which, rather than
creating vorticity with the giant vortex-pins set up here proposed, we simply numerically imprint a given initial
number of vortices uniformly at random position onto the same harmonically trapped condensate. We obtain essentially the following results: the polarization approximately retains its initial
value $P=0$ (see plot (f) in FIG.~\ref{fig4}), and a reasonable fit is obtained using Eq.~(\ref{logistic}). As opposed to the polarized cases, we find non-negative rates. However, we see in the comparison shown in FIG.~\ref{fig5} that Eq.~(\ref{ntotdecay}) fits the curve better, showing that in this case the non-linear term should be $\propto N^4$ instead of $\propto N^2$ to account for the steeper decay. Similarly to the polarized cases, drifting has shown to be the main mechanism of vortex loss. Eq.~(\ref{logistic}) has fitted well the cases studied in \cite{Kwon2014,Stagg2015}, which differs to our $P=0$ case not in polarization but rather in the initial total number of vortices (in their case $\sim 60$ as opposed to $\sim 140$ in ours). Therefore, we attribute the strong non-linear effects and the departure from the quadratic nature (which is consistent with the `ultra-quantum' decay observed \cite{Walmsley2008,Baggaley2012} in superfluid helium, where system's finiteness is not an issue) to vortex mutual interaction in a system which is finite. The finiteness imposes a certain limit to the number of vortices which can be accommodated in the cloud. 

In summary, for both polarized case and the particular unpolarized case where the vortex density is high, our proposed modified decay rate equation, Eq.~(\ref{ntotdecay}) successfully describes the total vortex number evolution. Furthermore, it adequately applies to our particular dissipative simulations.

\section{Conclusion}

We have presented a new scheme for generating 2D quantum turbulence
in atomic condensates which allows control over the polarization of
the flow, equivalent to the net rotation of a turbulent ordinary
fluid. Using this experimentally feasible scheme, we have examined the decay of the turbulence
and the vortex interactions (vortex-antivortex creation and
annihilation) which take place in the condensate. These results modeling the decay of the number of vortices using a new model
equation that takes into account polarization and its time-dependence; we have found that, in this context, it offers more reasonable fits than the logistic equation proposed by \cite{Kwon2014}. The reason being it includes polarization in the description and also accounts for more non-linear (steeper than quadratic) effects that have proven to be important in the initial decay process. 

In short, the inclusion of the time-dependent polarization parameter to the rate equation of the vortex number seems to capture the complete dynamics. It can offer experimentalists a hint on the ratio of positive and negative vortices in a 2D system, which is still an experimental challenge nowadays.

\section*{Acknowledgments}

We thank G. W. Stagg for useful discussions and also W. G. Kwon and Y. Shin for insightful suggestion on the phenomenological model, particularly for proposing the inclusion of polarization in the vortex number rate equation as a non-linear term $\propto N^{+}N^{-}$. 

This research was financially supported by CAPES (PDSE Proc. number BEX 9637/14-1), CNPq, and FAPESP.
LG's work is supported by Fonds National de la Recherche, Luxembourg, Grant n.7745104. The N´ucleo de Apoio a Optica e Fotˆonica (NAPOF-USP) ´
is acknowledged for computational resources. This work made use of the facilities of N8 HPC Centre of Excellence, provided and funded by the N8 consortium and EPSRC (Grant No.EP/K000225/1).

\bibliography{bib}

\end{document}